\documentclass[12pt]{iopart}

\usepackage{graphicx}
\begin{document}

\title{Superconducting properties of nanocrystalline MgB$_2$}

\author{B Lorenz$^1$, O Perner$^2$, J Eckert$^3$ and C W Chu$^{1,4,5}$}

\address{$^1$ Department of Physics and TCSUH, University of Houston, Houston, TX 77204-5002}
\address{$^2$ IFW Dresden, Institute of Metallic Materials, Helmholtzstrasse 20, D-01069 Dresden, Germany}
\address{$^3$ FB 11 Material- und Geowissenschaften, FG Physikalische Metallkunde, TU Darmstadt, D-64287 Darmstadt, Germany}
\address{$^4$ Lawrence Berkeley National Laboratory, 1 Cyclotron Road, Berkeley, CA 94702}
\address{$^5$ Hong Kong University of Science and Technology, Hong Kong, China}
\ead{blorenz@uh.edu}

\begin{abstract}
Nanocrystalline MgB$_2$ produced by mechanical alloying has been
shown to exhibit enhanced superconducting properties such as
increased pinning and higher critical currents. However, the effects
of the synthesis process on the intrinsic superconducting properties
have not been addressed yet. We have investigated the
superconducting gap structure of nanocrystalline MgB$_2$ pellets
synthesized by high-energy ball milling employing specific heat
measurements. We found that the larger $\sigma$-gap decreased
whereas the smaller $\pi$-gap slightly increased in ball milled
MgB$_2$ as compared to bulk samples synthesized along the standard
routes. The data show that the ball milling process introduces
defects that enhance the interband scattering similar to irradiation
with neutrons. The reduction of the $\sigma$-gap explains the lower
superconducting transition temperature of 33 K.
\end{abstract}

%Uncomment for PACS numbers title message
%\pacs{00.00, 20.00, 42.10}
% Keywords required only for MST, PB, PMB, PM, JOA, JOB?
%\vspace{2pc}
%\noindent{\it Keywords}: Article preparation, IOP journals
% Uncomment for Submitted to journal title message
%\submitto{\JPA}
% Comment out if separate title page not required
\maketitle

\section{Introduction}
Since the discovery of superconductivity at 40 K in MgB$_2$ \cite{1}
many attempts have been reported to improve the material properties
that are important for potential applications. Unlike the high-T$_c$
cuprate superconductors where the current carrying capacity is
limited by weak links across grain boundaries, the critical current
in MgB$_2$ is mainly determined by its flux pinning properties.
Various investigators have therefore focused their attention onto
the controlled increase of the number of pinning centers by
introducing disorder and defects using, for example, irradiation
techniques \cite{2} or the implementation of nanometer-sized
particles of Si, C, SiC, TiB, or YB$_4$ into polycrystalline MgB$_2$
\cite{3}. An increase of the upper critical field was in fact
observed in bulk samples as well as in thin films showing a high
degree of disorder in the boron planes \cite{4}. Alternatively, the
pinning properties of MgB$_2$ can be improved by increasing the
density of grain boundaries and/or bulk defects by reducing the
average grain size into the nanometer range. Nanocrystalline powders
can be produced by mechanical alloying the starting materials (Mg
and B) using high-energy ball milling \cite{5}. The partial reaction
of Mg and B to MgB$_2$ during the milling process is completed by
subsequent hot pressing resulting in very dense ceramic pellets with
distinctly enhanced pinning in the superconducting state. A
combination of both routes, the addition of oxide particles into a
nanocrystalline MgB$_2$ matrix has led to a significant improvement
of the extrinsic superconducting properties \cite{6}.

The introduction of disorder and defects, however, has also an
effect on other relevant parameters of the superconducting state.
First of all, in many instances the critical temperature, T$_c$, is
dramatically reduced (by up to 20 \% of the maximal T$_c$ $\simeq$
40 K) as defects and disorder are implemented to enhance H$_{c2}$.
This defect-induced decrease of T$_c$ was discussed in an early
review on MgB$_2$ \cite{7} in analogy to the reduction of T$_c$ in
ion-irradiated superconducting Nb-Ge films \cite{8}. Of more
fundamental interest, however, is the influence of impurities,
disorder, and defects on the superconducting gap structure of
MgB$_2$. The possible existence of multiple superconducting gaps was
first suggested by Liu et al. from a first-principle calculation of
the band structure and the electron-phonon coupling in MgB$_2$
\cite{9} and later confirmed by a number of experimental
investigations, preferentially heat capacity measurements \cite{10}
and tunneling spectroscopy \cite{11}. The two superconducting gaps
are correlated via interband scattering and they close at the same
critical temperature. The existence of two types of supercarriers
and two gaps of different magnitude is the origin of the very
peculiar temperature dependence of the heat capacity, C$_p$(T), in
the superconducting state which deviates from the typical
T-dependence of C$_p$ following from the BCS theory \cite{12}. The
larger gap was assigned to the  $\sigma$-band the carriers of which
couple strongly to the E$_{2g}$ phonons in the boron planes whereas
the smaller gap appears in the  $\pi$-band and its carriers couple
only weakly to phonons. A direct experimental proof for this
assignment was given recently \cite{13}. From a theoretical point of
view the problem of multi-band superconductivity was considered
already several decades ago and it was found that the two
superconducting gaps close at the same T$_c$ as a consequence of
interband coupling between electrons in different bands mediated by
phonons \cite{14}.

It is the main objective of this work to investigate the effects of
the small grain size of nanocrystalline MgB$_2$ and the impurities
and defects introduced by the high-energy ball milling procedure on
the intrinsic superconducting properties and, in particular, on the
superconducting gaps of MgB$_2$. Magnetic, electrical transport, and
heat capacity experiments are conducted and the values for the
superconducting gaps are extracted from the temperature dependence
of the specific heat. We show that the larger $\sigma$-gap is
reduced by more than 40 $\%$ in nanocrystalline MgB$_2$ as compared
with samples prepared using standard synthesis procedures.

The experimental setup is described in section 2 and the results are
presented in section 3. The conclusions are presented and discussed
in the last section.

\section{Experimental techniques}
Nanocrystalline MgB$_2$ was prepared by high-energy ball milling of
the precursor materials Mg (99.8 $\%$, particle size 250 $\mu$m) and
B (99.9 $\%$, particle size 1 $\mu$m) as described in more detail
elsewhere \cite{5}. The milling was conducted in Ar atmosphere using
jar and balls made from tungsten carbide (WC). After 20 hours of
milling time Mg and B partially reacted (cold alloying) to form
MgB$_2$. The x-ray spectrum shows already some peaks assigned to
MgB$_2$ together with diffraction peaks from Mg and WC. The reaction
was completed by hot uniaxial pressing at 973 K / 640 MPa for 10
minutes. This treatment resulted in a complete conversion of Mg and
B to MgB$_2$ with only minor traces of impurity phases left (3 $\%$
MgO and 0.3 $\%$ WC, Fig. 1 and \cite{5}). The coherent scattering
length for the MgB$_2$ was estimated from the x-ray spectra as 30 nm
and the MgB$_2$ particle size (from SEM) is 40 to 100 nm.

The physical properties and the superconducting transition are
characterized by ac magnetic susceptibility, resistivity,
thermoelectric power, and specific heat measurements. The ac
susceptibility was determined by measuring the mutual inductance of
a dual coil system attached to the sample employing the LR700 Mutual
Inductance/Resistance Bridge (Linear Research). The resistivity was
measured in four-lead configuration with the same device. For the
thermoelectric power experiments we have used a home made
high-resolution ac measurement technique. The heat capacity was
determined by a relaxation technique using the Physical Property
Measurement System (Quantum Design).

\section{Results and discussion}
The electrical resistivity (Fig. 2) displays the characteristic
temperature dependence of MgB$_2$ with a low ratio of the residual
(R just above T$_c$) and room temperature resistance, RRR=1.38,
which is typical for MgB$_2$ samples with a high concentration of
defects or impurities \cite{25}. Enhanced scattering at the grain
boundaries should also contribute to an increase of the residual
resistance and lower the RRR-values. The inset of Fig. 2 shows the
ac susceptibility near the superconducting phase transition. The
T$_c$=33 K defined by the midpoint of the drop of $\chi_{ac}$ is
significantly lower than the 39 K known for bulk MgB$_2$. The width
of the transition of only 0.6 K indicates good sample uniformity
which is important for the discussion of the thermodynamic
properties.

The thermoelectric power, S(T), is less sensitive to grain
boundaries and reflects the intrinsic material properties better
than the resistivity. The thermoelectric power of the
nanocrystalline MgB$_2$ (Fig. 3) is well comparable with data
obtained from "standard" MgB$_2$ \cite{15,16}. The inset of Fig. 3
expands the temperature range close to T$_c$ and reveals the sharp
drop of S(T) to zero at the superconducting transition. The change
of slope above 200 K and the high-temperature value of about 9
$\mu$V/K is very similar to the properties of bulk samples. The
major difference is the lower T$_c$ value that indicates an effect
of the grain size and the ball milling procedure on the extrinsic
superconducting properties.

The superconducting gaps of MgB$_2$ can be determined from
thermodynamic quantities such as the specific heat. It was
demonstrated that the particular two-gap structure is responsible
for the peculiarities of the heat capacity observed at low T and
close to T$_c$ \cite{10}. For bulk MgB$_2$ the thermodynamic
description based on the "$\alpha$-model" \cite{17} was very
successful and the two superconducting gaps extracted from the
specific heat data were found in good agreement with data from
tunneling spectroscopy \cite{11}. Within this model the
thermodynamic properties of a two-gap (or two-band) superconductor
are calculated based on a superposition of contributions from the
carriers of the two contributing bands assuming a BCS-like
temperature dependence \cite{12} for each of the two gaps of
different magnitude but with the same critical temperature, T$_c$.
The validity of the $\alpha$-model as a first approximation to
determine the superconducting gaps of MgB$_2$ has been demonstrated
recently \cite{18}. It was successfully used to determine the change
of the gap structure in neutron irradiated \cite{19} or Al
substituted MgB$_2$ \cite{20}.

The specific heat was measured for nano crystalline and bulk
MgB$_2$. The lattice contribution was subtracted from the data by
extrapolating the high-temperature specific heat to zero T and
maintaining the entropy balance of the superconducting state. The
electronic specific heat as a function of the reduced temperature is
shown for the nano crystalline and bulk MgB$_2$ samples as full and
open symbols in Fig. 4. The significant difference in the
temperature dependence indicates the intrinsically different
superconducting gap structure of both samples. Using the
$\alpha$-model to fit the experimental data for bulk MgB$_2$ the
estimated gap values of $\Delta_\sigma$=6.9 meV and
$\Delta_\pi$=1.93 meV are in excellent agreement with previous
reports from specific heat and tunneling spectroscopy. For nano
crystalline MgB$_2$, however, the fit of the two-gap $\alpha$-model
leads to a $\sigma$-band gap that is considerably reduced
($\Delta_\sigma$=3.77 meV) whereas the $\pi$-gap slightly increased
to $\Delta_\pi$=1.99 meV. In addition, there is a re-distribution of
weight in that almost 2/3 of the weight contributing to the heat
capacity of the nano crystalline sample results from the
$\sigma$-band as compared to a 1:1 distribution in bulk MgB2.

The current data show a similar change of the superconducting gap
values as has been reported for neutron-irradiated and
Al-substituted MgB$_2$ samples \cite{19,20}. In both cases the
larger $\sigma$-gap is reduced by either introducing defects or by
replacing Mg with Al whereas the effect on the smaller $\pi$-gap is
relativelt small. A correlation has been established between T$_c$
and the value of $\Delta_\sigma$. This correlation appears to be
universal in that T$_c$ decreases linearly with $\Delta_\sigma$ and
it holds well for most of the substituted and irradiated samples
\cite{19,20,21}. Our values for T$_c$ (33 K) and $\Delta_\sigma$
(3.8 meV) of nanocrystalline MgB$_2$ are consistent with the
reported data and with the results of a two-band Eliashberg theory
calculation correlating the T$_c$ of Mg$_{1-x}$Al$_x$B$_2$ with the
ratio of the two superconducting gaps, $\Delta_\pi$/$\Delta_\sigma$
\cite{22}. The origin of the $\sigma$-gap suppression in the
ball-milled nanocrystalline MgB$_2$ samples has yet to be resolved.
Two physical mechanisms, the band filling effect and the increase of
interband scattering, and their effects on the gap structure of Al
or carbon substituted MgB$_2$ have recently been proposed \cite{23}.
Whereas both effects lead to a decrease of the $\sigma$-gap their
influence on the smaller $\pi$-gap is opposite. The change of band
filling decreases the $\pi$-gap but the interband scattering tends
to increase the smaller gap. Both mechanisms may actually compete
with one another explaining the opposite tendency of the measured
$\pi$-gaps in Al and carbon substituted MgB$_2$. The dominating
physical mechanism in substituted MgB$_2$ is still a matter of
controversy \cite{24}. For our nanocrystalline samples of MgB$_2$ it
appears conceivable that the filling of the bands is not changed by
the synthesis procedure since there is no apparent doping that could
affect the carrier density. The introduction of defects or vacancies
as a consequence of stresses induced during the high-energy ball
milling procedure and of a minor impurity phase (MgO and WC
according to the x-ray spectra) is more likely to be the cause of
the reduction of T$_c$ and the $\sigma$-gap. The latter assumption
is supported by the small increase of the $\pi$-gap and our results
are similar to the case of defects introduced by neutron irradiation
\cite{19}. The current data reveal another interesting example where
enhanced interband scattering in MgB$_2$ dominates and causes the
drop of T$_c$ as well as the characteristic change of the two
superconducting gaps.

\section{Summary}
Nanocrystalline MgB$_2$ has been synthesized by high energy ball
milling of MgB$_2$ powder. The samples show an enhanced critical
current density due to improved pinning properties and a reduced
T$_c$. The intrinsic superconducting gap structure was revealed by
heat capacity measurements. The data can be described by the
two-band $\alpha$-model. The larger $\sigma$-gap is reduced but the
smaller $\pi$-gap slightly increases as a result of the synthesis
process. The results show that the interband scattering is enhanced
due to defects introduced by the ball milling process.

\ack {This work is supported in part by NSF Grant No. DMR-9804325,
the T.L.L. Temple Foundation, the John J. and Rebecca Moores
Endowment, and the State of Texas through the TCSUH at the
University of Houston and at Lawrence Berkeley Laboratory by the
Director, Office of Energy Research, Office of Basic Energy
Sciences, Division of Materials Sciences of the U.S. Department of
Energy under Contract No. DE-AC03-76SF00098.}

\newpage

\begin{figure}[tbp]
%\begin{center}
%\includegraphics[angle=-90,width=4in]{Fig1.eps}
%\end{center}
\caption{X-ray spectrum of nanocrystalline MgB$_2$ synthesized by
high-energy ball milling and subsequent heat treatment. Small peaks
of a minor impurity phase of WC are indicated in the spectrum.}
%\end{figure}

%\begin{figure}[tbp]
%\begin{center}
%\includegraphics[angle=-90, width=4in]{Fig2.eps}
%\end{center}
\caption{Resistivity and ac-susceptibility (inset) of
nanocrystalline MgB$_2$. The sharp transitions at T$_c$=33 K prove
the uniformity of the sample.}
%\end{figure}

%\begin{figure}[tbp]
%\begin{center}
%\includegraphics[angle=-90, width=4in]{Fig3.eps}
%\end{center}
\caption{Thermoelectric power of nanocrystalline MgB$_2$. Its
temperature dependence and magnitude are comparable to bulk
MgB$_2$.}
%\end{figure}

%\begin{figure}[tbp]
%\begin{center}
%\includegraphics[angle=-90, width=4in]{Fig4.eps}
%\end{center}
\caption{Electronic heat capacity for nanocrystalline (closed
symbols) and bulk (open symbols) MgB$_2$. The lines show the fit to
the $\alpha$-model. Note that the entropy balance is fulfilled in
both data sets.}
\end{figure}

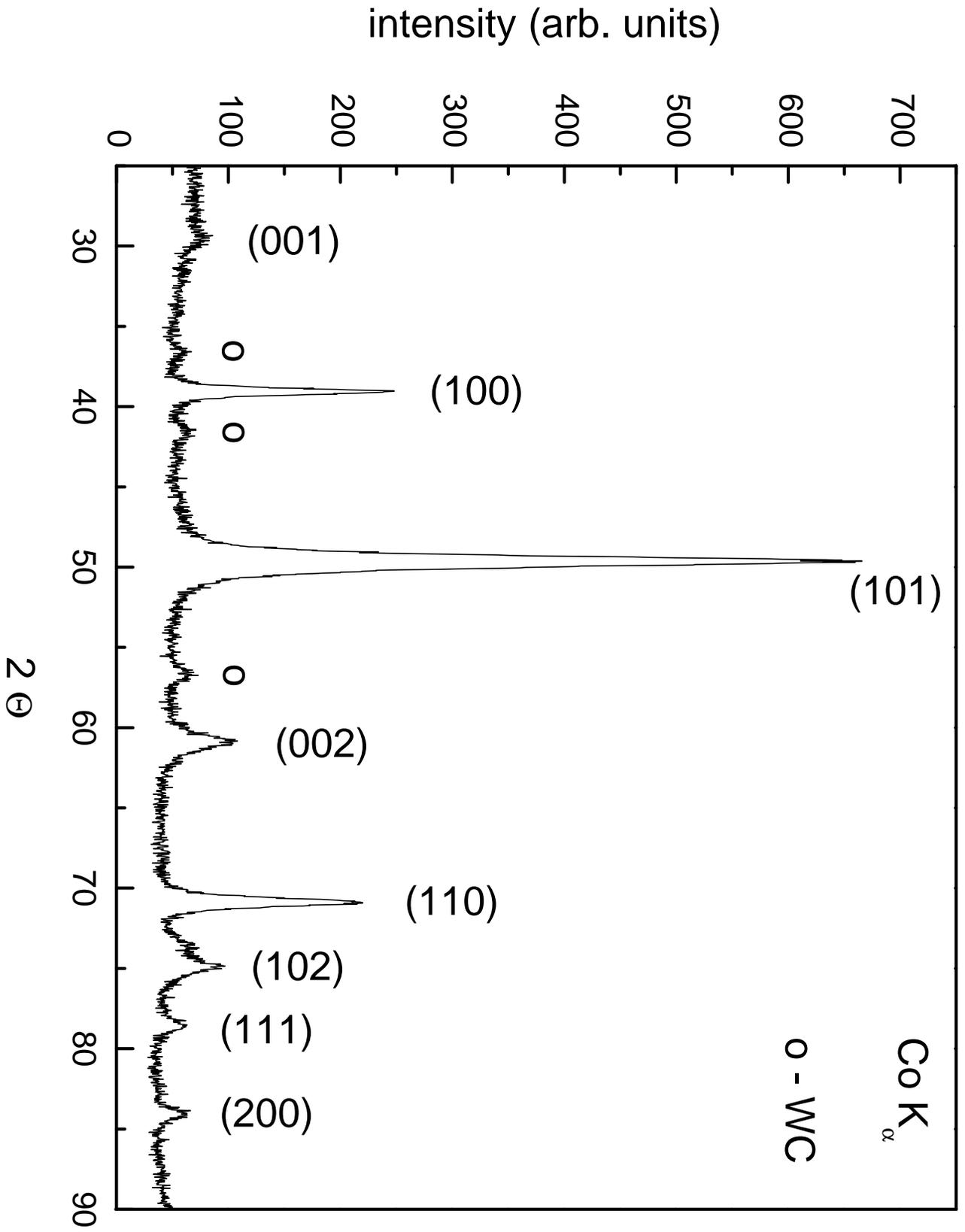
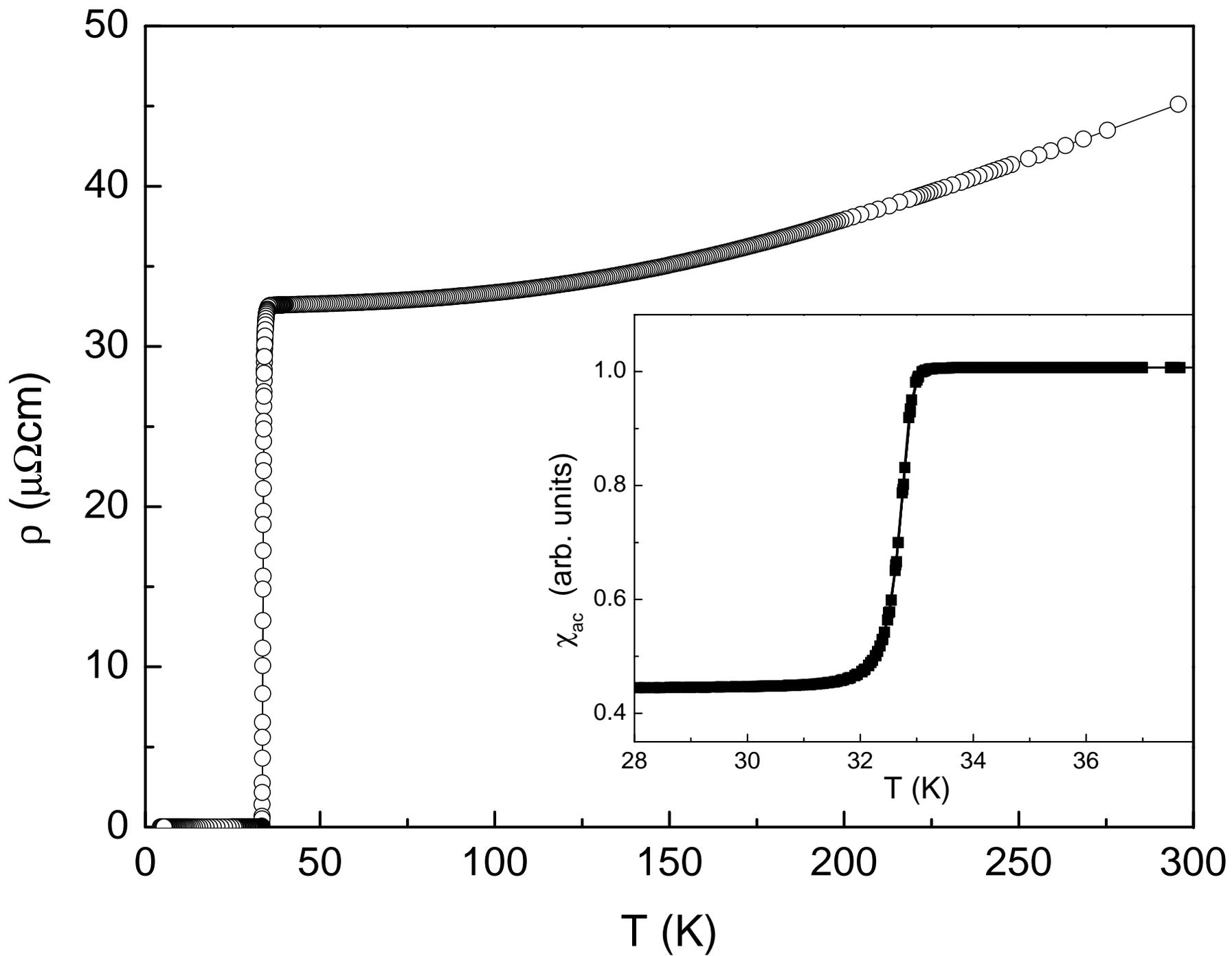
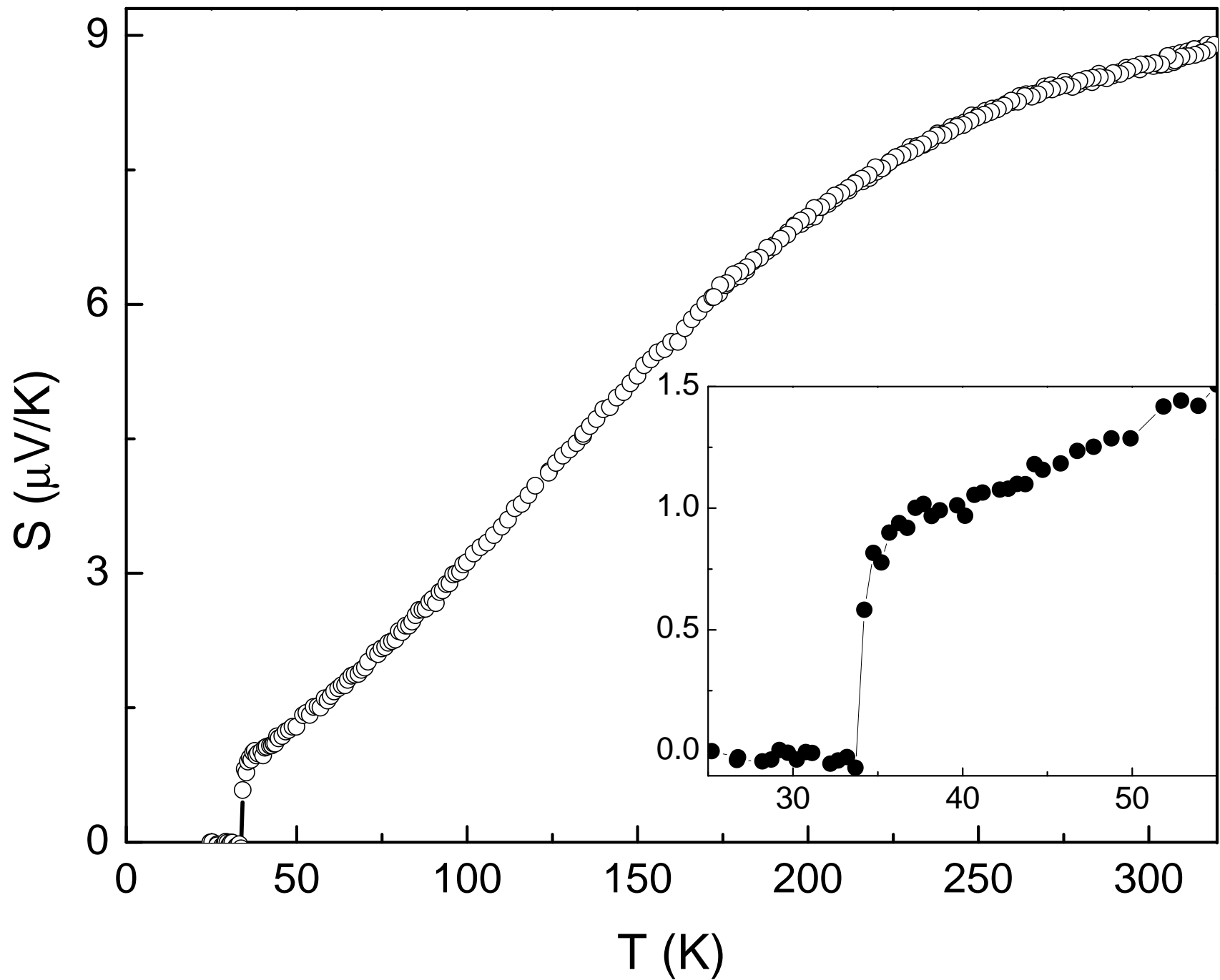
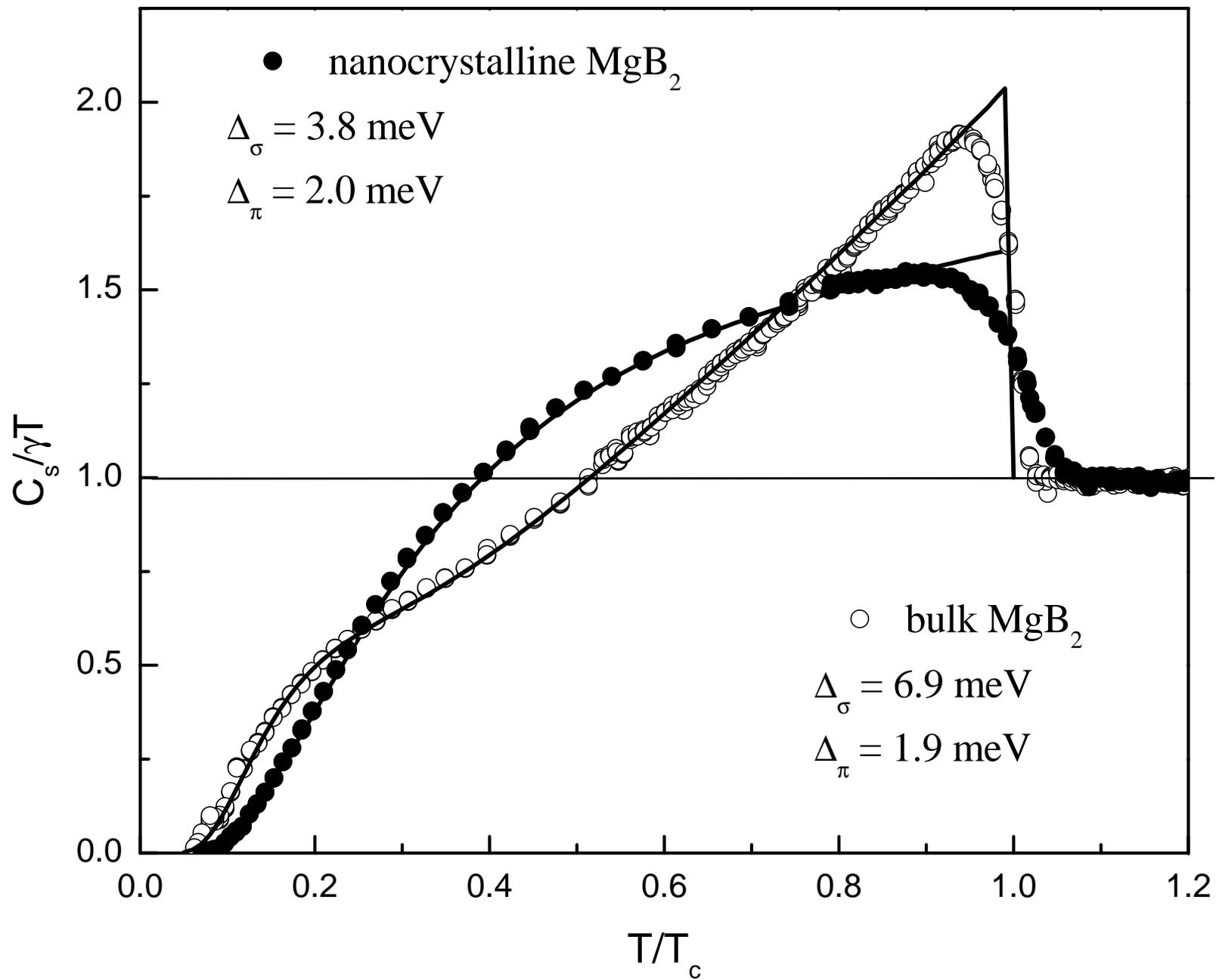
\end{document}